\documentclass[conference]{IEEEtran}

\usepackage{csquotes}
\usepackage{enumitem}
\usepackage{array}
\usepackage{tikz}
\usepackage{todonotes}
\usepackage{subcaption}
\usepackage{xspace}
\usepackage{multirow}
\usepackage[table,xcdraw]{xcolor}
\usepackage{listings}
\usepackage[framemethod=TikZ]{mdframed}
\usepackage{pifont}
\usepackage{booktabs}
\usepackage[most]{tcolorbox}
\usepackage{cleveref}
\tcbuselibrary{breakable}
\usepackage{framed}
\usepackage{placeins}
\usepackage{tabularx}
\usepackage{pifont}
\usepackage{xcolor}
\usepackage{soulutf8}
\definecolor{dgreen}{RGB}{1,50,32}

\newcommand{\stpatool}{SAMD\xspace}

\newcommand{\fullcircle}[2][black]{
    \begin{tikzpicture}
        \node[circle, draw=#1, fill=#1, minimum size=#2] {};
    \end{tikzpicture}
}

\newcommand{\halfcircle}[2][black]{
    \begin{tikzpicture}
        \node[circle, draw=#1, path picture={\fill[#1] (path picture bounding box.north west) rectangle (path picture bounding box.south);}, minimum size=#2] {};
    \end{tikzpicture}
}

\newcommand{\emptycircle}[2][black]{
    \begin{tikzpicture}
        \node[circle, draw=#1, fill=none, minimum size=#2] {};
    \end{tikzpicture}
}

\begin{document}




\title{\stpatool: A Tool for Identifying False Data Injection Scenarios in AI/ML-enabled Medical Devices}

\author{\IEEEauthorblockN{Mohammadreza Hallajiyan\textsuperscript{$\dagger$}, Xueren Ge\textsuperscript{$\ddagger$}, Athish Pranav Dharmalingam\textsuperscript{$\uparrow$}, Gargi Mitra\textsuperscript{$\dagger$},\\Shahrear Iqbal\textsuperscript{$\ast$}, Homa Alemzadeh\textsuperscript{$\ddagger$}, Karthik Pattabiraman\textsuperscript{$\dagger$}}

\IEEEauthorblockA{
\textsuperscript{$\dagger$}University of British Columbia, \textsuperscript{$\ddagger$}University of Virginia, \textsuperscript{$\uparrow$}IIT Madras, \textsuperscript{$\ast$}National Research Council Canada\\
Email: hallaj@ece.ubc.ca, zar8jw@virginia.edu, athishanna@gmail.com, gargi@ece.ubc.ca,\\ shahrear.iqbal@nrc-cnrc.gc.ca, ha4d@virginia.edu, karthikp@ece.ubc.ca}

}
\maketitle




\begin{abstract}



The growing integration of artificial intelligence (AI) and machine learning (ML) in medical systems requires effective measures to address emerging security risks. One such risk is that of adversaries introducing false data through vulnerable system components during inference, causing misdiagnosis and wrong treatments. These risks are challenging to anticipate and address in the design phase, as the system assembly partially occurs during actual use by end users. To address this concern, we introduce \stpatool, an automated tool for performing System Theoretic Process Analysis for Security (STPA-Sec) on AI/ML-enabled medical devices during the design phase. \stpatool models the medical system as a control structure, treating all system components as potential points for injecting false data into the ML engine. It leverages state-of-the-art vulnerability databases and Large Language Models (LLMs) to automate vulnerability discovery and generate a list of potential attack scenarios. We demonstrate \stpatool's effectiveness through case studies on five FDA-cleared medical devices, showcasing its ability to identify vulnerable points and potential attack paths. 
We find that \stpatool has 100\% precision in identifying target device technologies in the case studies' documents, retrieves the known vulnerabilities linked to them (with 63.2\% precision), and generates highly relevant attack scenarios on the ML model, including detailed steps that an adversary might take (with 95.3\% accuracy, and the highest time taken being 191.64s).
\end{abstract}


\section{Introduction}\label{sec:intro}

There is a growing trend to incorporate artificial intelligence (AI) and machine learning (ML) into medical applications \cite{tariq3principles} to automate remote patient monitoring, enhance diagnosis and treatment, and improve healthcare accessibility in remote locations. Over 1000 such devices are registered with the U.S. FDA as of 2024~\cite{fdaml}, including  in life-critical scenarios, such as live surgical assistance~\cite{nuvasive} and generation of treatment plans~\cite{ondrop}. Therefore, mispredictions by the ML algorithms may lead to health complications and even death of the patient. 
We focus on mispredictions due to security attacks, \textit{specifically inference-time false (input) data injection attacks}~\cite{newaz2020adversarial,wang2022threats}. 

Detecting and preventing false data injection attacks in AI/ML-enabled medical systems is challenging. Malicious inputs are difficult to detect due to the complexity and often unexplainable nature of ML models~\cite{arrieta2020explainable}, along with variations in patients' physiological characteristics.
Preventing these attacks in AI/ML-enabled medical systems is equally challenging because of the large and complex attack surface, involving multiple interconnected devices. 

ML-enabled medical devices differ fundamentally from other safety-critical domains such as industrial control, automotive systems, or conventional home IoT. In those settings, a single manufacturer typically designs and integrates both the device and its associated cloud software, enabling uniform security assumptions. In contrast, ML-enabled medical systems rely on heterogeneous, multi-vendor peripheral devices that supply inputs to the ML model. The final integration of the ML software (standalone or embedded in an IoT device) with these peripherals is often performed post-deployment by patients or healthcare providers, who lack security expertise. Moreover, ML-enabled applications are frequently designed to interoperate with multiple brands of each peripheral (e.g., glucose meters), each with different vulnerabilities. Therefore, even if the AI/ML-enabled software is secure, the post-deployment security risks are difficult to predict at design time. Although vulnerabilities in peripheral devices may fall outside the ML developer’s direct implementation scope, the ML-enabled system ultimately generates the clinical decision. Consequently, harm caused by adversarially manipulated inputs may still make the manufacturer liable for failing to anticipate or mitigate upstream risks.

Existing efforts to secure AI/ML-enabled systems focus on enhancing the resilience of medical ML models~\cite{qayyum2020secure}, identifying vulnerabilities within the medical software stack (e.g., an ML-enabled patient breathing rate monitor)\cite{mandalari2023protected} and its associated datastore that integrates all IoT device inputs\cite{kafle2019study}, its host IoT device\cite{kustosch2025patching}, or identify hazards in cloud-connected IoT devices~\cite{zhou2019discovering}. However, none of them focus on identifying known vulnerabilities in peripheral devices (e.g., the camera live-feeding the breathing monitor) and assessing whether their exploitation might lead to hazardous ML model outputs.
Anticipating these scenarios would enable both device manufacturers and independent security analysts to foresee and mitigate post-deployment security risks early~\cite{mitra2025misc}. 

\par System Theoretic Process Analysis (STPA) is a widely used methodology that facilitates the early identification of potential safety hazards in a system using a control-theoretic perspective~\cite{young2013systems}. STPA-Sec, a variant of STPA, extends the focus to security-related threats~\cite{fleming2023introduction}. 
We developed \textbf{\stpatool{}} (STPA-Sec for AI/ML-enabled Medical Devices), a tool\footnote{We will make the tool open-source after paper acceptance} to identify system-wide security risks due to false data injection in AI/ML-enabled medical systems. \stpatool{} adopts STPA-Sec, to identify how an adversary can exploit security vulnerabilities in interconnected components to cause the system to malfunction, potentially leading to a safety hazard. 

\stpatool{} leverages Named Entity Recognition (NER) techniques, vulnerability databases, and the reasoning ability of Large Language Models (LLMs) to automatically detect \emph{all possible ways} an adversary can exploit known vulnerabilities in peripheral devices to inject malicious data into an AI/ML-enabled device, leading to either misdiagnoses or incorrect treatment plans, eventually harming the patients. 
\emph{To the best of our knowledge, \stpatool is the first automated tool to assist STPA-Sec analysis of AI/ML-enabled medical devices.}

\par There are two groups of people who can benefit from \stpatool's output, i.e., identified attack paths \textit{(1) device manufacturers} who will benefit by using \stpatool{} preferably at the design phase\textemdash which is a key advantage of \stpatool{}\textemdash and identify potential attack paths to implement countermeasures ideally before commercialization of the device or issue advisories to end users and IT personnel of healthcare delivery organizations on which peripheral devices to avoid and how to set up a secure deployment environment, and 
\textit{(2) independent security analysts}, who would use publicly available information on AI/ML-enabled medical devices to perform security analysis using \stpatool and come up with a set of security recommendations.

\textbf{Contributions:} Our contributions are as follows.
\begin{itemize}[leftmargin=*]
    \item We devise \stpatool{}, a tool to assist security analysts with the identification of potential false data injection scenarios and their impact in AI/ML-enabled devices. 
    \stpatool{} can automatically (1) identify potential target device technologies and (2) associated vulnerabilities from publicly available resources, and (3) generate potential attack scenarios with specific steps for injecting false data into the ML model of a target AI/ML-enabled device.
    \item We demonstrate the efficacy of \stpatool{} by implementing it using three well-known LLM models (GPT-4, GPT-4o and Llama 3) and applying it to the analysis of five real-world FDA-cleared AI/ML-enabled devices (d-Nav~\cite{dnav}, ABMD~\cite{abmd}), IDx-DR v2.3 ~\cite{idx}, KIDScore D3~\cite{Kidscore}, Oxehealth Vital Signs~\cite{oxehealth}) from various medical specialties. 
    \item We evaluate \stpatool by manual review of the results from its components and by assessing the generated attack steps using a LLM-as-a-judge method. The results show that \stpatool is highly effective in identifying target device technologies in the input documents (with 100\% precision), successfully retrieves known vulnerabilities linked to them (with 63.2\% average precision), and generates highly relevant attack scenarios on the ML model, including detailed steps that an adversary might take (with 95.3\% average accuracy, and the highest time taken being 191.64s.).
\end{itemize}

\section{Background and Related Work}
\label{background}
\textbf{STPA.} System Theoretic Process Analysis (STPA) is a well-established methodology for identifying potential hazards in complex systems~\cite{leveson2011engineering}. 
By identifying the underlying causes of hazards\textemdash across the layers of system control structure, referred to as causal scenarios\textemdash STPA has been proven as a valuable method for ensuring system safety~\cite{leveson2013stpa}. 

\textbf{STPA-Sec.} Building on STPA, STPA-Sec\cite{fleming2023introduction} identifies  causes of hazards due to security threats by introducing the concept of adversarial thinking into the analysis process \cite{young2013systems, young2017system}, and has been used in domains such as Autonomous Vehicles (AVs) \cite{sabaliauskaite2018integrating, sharma2019safety} and medical devices \cite{hallajiyan2023sam,mitra2025misc}.
STPA-Sec offers a structured framework for linking security vulnerabilities to hazardous system states using the following steps\cite{fleming2023introduction}: \textbf{(i)} identifying the unacceptable losses, as defined by the system's functional goals, and the undesirable events; \textbf{(ii)} identifying the possible hazards that might lead to unacceptable losses; \textbf{(iii)} identifying control actions in the system; \textbf{(iv)} defining the circumstances in which a particular control action might be unsafe; and, \textbf{(v)} Identifying the set of constraints that should be applied to limit the possibility of a hazardous scenario leading to an unacceptable loss.
\textbf{Tools for STPA and STPA-Sec.} Several tools have been developed to facilitate STPA and STPA-Sec in different application domains. To determine their suitability for STPA-Sec on ML-enabled medical systems, Table \ref{tab:related} summarizes their main features, including: (1) \emph{Focus} - whether they address security attacks on ML-enabled systems, (2) \emph{Application Domain} - to what domain they are applied, and (3) \emph{Automation Level} - whether they automate the process of causal scenario generation. 
A-STPA \cite{abdulkhaleq2014open} and its enhanced version, XSTAMPP \cite{abdulkhaleq2015xstampp} help a user link unsafe control actions to identified safety hazards and provide graphical aids for control structure creation, but require manual identification of causal scenarios. SafetyHAT \cite{becker2014transportation} is a customized tool for the transportation sector, which offers a graphical interface, data management capabilities, and transportation-specific guidewords for identifying unsafe control actions and causal scenarios, but still requires manual identification of causal scenarios. WebSTAMP \cite{souza2019webstamp} is a web application designed for STPA and STPA-Sec, that aims to provide a structured, automated, and comprehensive analysis. The tool offers a list of comprehensive control actions and guiding questions to help user identify hazardous control actions and their causal scenarios. 
While WebSTAMP offers a systematic approach, its ability to fully automate the process, especially in the context of external threats in STPA-Sec, remains limited. \textbf{(4)} SOT \cite{pereira2019stamp} helps systems engineers conduct safety and security analyses, using stored knowledge from previous analyses to identify causal scenarios. While beneficial, SOT still relies on analysts to devise new scenarios manually or update the proposed ones, and given the continuous emergence of vulnerabilities, it may struggle to remain current.
In summary, A-STPA, XSTAMPP, and SafetyHAT focus only on safety concerns caused by device failures, not malicious data injection attacks. WebSTAMP and SOT address security concerns but rely heavily on users' knowledge of security vulnerabilities and manual effort. 



\begin{table}[t]
\scalebox{1}{
\setlength{\tabcolsep}{1pt}  
\begin{tabular}{|m{2.5cm}|c|>{\centering\arraybackslash}m{2.5cm}|>
{\centering\arraybackslash}m{1.5cm}|}
    \hline \centering
    \footnotesize
    \textbf{Name} & \centering \textbf{Focus} & \textbf{Application Domain} & \textbf{\hspace{5mm}Automation Level \newline (Attack Scenarios)} \tabularnewline
    \hline 
     \centering A-STPA~\cite{abdulkhaleq2014open}, XSTAMPP~\cite{abdulkhaleq2015xstampp}& Safety & \centering General Purpose & \halfcircle[black]{2.5ex}\\
     \hline
     \centering SafetyHAT~\cite{becker2014transportation} & Safety & \centering Transportation & \halfcircle[black]{2.5ex}\\
     \hline
     \centering WebSTAMP~\cite{souza2019webstamp} & Safety/Security & \centering Healthcare, Transportation, Chemical Industry & \halfcircle[black]{2.5ex}\\
     \hline
     \centering SOT~\cite{pereira2019stamp} & Safety/Security & \centering Aircraft Systems &  \halfcircle[black]{2.5ex}\\
     \hline
\end{tabular}}
\captionsetup{justification=centering}
\caption{State-of-the-Art STPA/STPA-Sec tools.\\ (\fullcircle[black]{1ex}: Fully-automated, \halfcircle[black]{1ex}: Semi-automated, \emptycircle[black]{1ex}: Manual)}
\label{tab:related}
\setlength{\tabcolsep}{6pt}
\vspace{-1em}
\end{table}


\textbf{Automation of STPA.} Recent work has attempted to automate STPA using LLMs. 
Qi et al.~\cite{qi2023safety} utilized ChatGPT to conduct a case study of STPA on the automatic emergency brake and demand side management systems. The study underscores the importance of expert intervention and proper prompt engineering to yield more relevant outcomes. 
Focusing on autonomous driving, Nouri et al. \cite{nouri2024welcome,nouri2024engineering} developed a carefully engineered prompt to be used in LLMs for extracting safety specifications. However, none of existing tools consider STPA-Sec, which requires considering adversarial scenarios where the causes are actions initiated by attackers. 
In contrast to these papers, \stpatool gathers vulnerabilities from state-of-the-art databases, uses LLMs, and incorporates adversarial attack and medical domain knowledge to provide a comprehensive list of potential causal scenarios (also referred to as \textit{attack paths} in this work). 

\section{Motivation and Challenges}\label{motivation}

\textbf{Motivating Example: }
 We present a Blood Glucose Management System (BGMS) as an example AI/ML-enabled medical system and highlight the security risks posed by vulnerabilities inherent in the system's interconnected structure. 
A BGMS enables patients to monitor and regulate their blood glucose levels by measuring current glucose levels and estimating the insulin dosage.

\begin{figure}[t]
    \centering
    \includegraphics[scale=0.45]{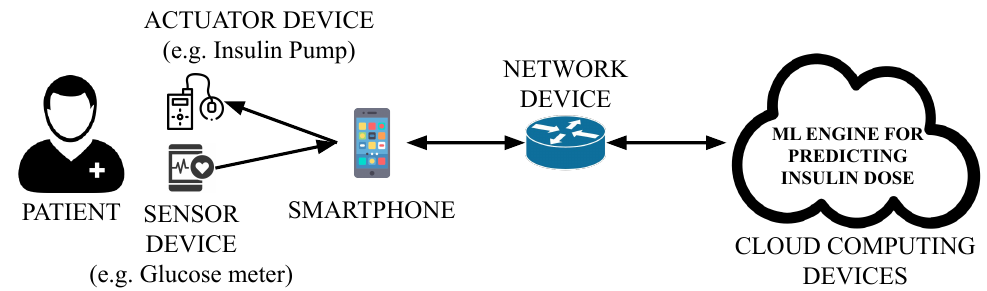}
    \caption{An example of BGMS including the peripheral devices and their interactions (adopted from~\cite{elnawawy2024}).}
    \vspace{-1em}
    \label{fig:BGMS_Example}
\end{figure}

Fig. \ref{fig:BGMS_Example} illustrates an example BGMS (adopted from~\cite{elnawawy2024}). 
In this system, a sensor device (e.g., a glucose meter) periodically measures the patient's blood glucose levels (e.g., every five minutes) and transmits them to an interface device, such as a mobile phone running an app (e.g., a blood glucose management app), where the patient’s data is displayed and stored. The glucose readings are then sent to a cloud-based ML engine, which uses the patient’s physiological characteristics to predict the required insulin dose. The predicted dose is sent back to the interface device, which either forwards the data to an automatic insulin pump for administration or displays it for the patient to inject the insulin manually.


\par Across this end-to-end setup, there are multiple points of vulnerability where attackers could manipulate data. The literature documents numerous real-world attacks~\cite{zhou2022design}, such as intercepting Bluetooth communication~\cite{antonioli2022blurtooth} between the glucose meter and mobile phone to alter data.  Vulnerabilities can also emerge in the mobile phone~\cite{senanayake2023android}, the network connection between the phone and the cloud \cite{nazir2021survey}, or even in the glucose meter or insulin pump, which may be susceptible to electromagnetic interference \cite{mortazavi2014electromagnetic}. Given the interconnected nature of the system, compromised data can flow into the ML model, resulting in incorrect insulin dose predictions.

\textbf{Challenges}: For medical device manufacturers, Steps i--iii of STPA-Sec (mentioned in \S{\ref{background}}) are tractable because they rely on domain knowledge of system functionality and clinical workflows. In contrast, Steps iv--v involve the following challenges, addressing which require cybersecurity expertise:

\textbf{First,} AI/ML-enabled medical systems include multiple interconnected devices, each with potential vulnerabilities in various technological layers (e.g., communication protocols, third-party software, operating systems, firmware). Identifying all these vulnerabilities is challenging, as it requires targeted data retrieval from vulnerability databases to gather information about vulnerabilities in the ML model, vulnerabilities in all compatible peripheral devices, and assessing their relevance to the medical device.  


\textbf{Second,} understanding how each vulnerability can cause the ML model to mispredict and affect patients requires expertise in ML vulnerabilities, system security, and medicine. This requires systematic reasoning about adversarial behaviors, attack vectors, and cross-component interactions. This analysis is important, but non-trivial, particularly given the limited cybersecurity expertise among medical ML developers. 

\textbf{Third,} despite the increasing use of LLMs to assist STPA, significant concerns remain about LLM hallucinations. 
We need to mitigate hallucinations to get trustworthy results~\cite{perkovic2024hallucinations}. 


\section{Approach: \stpatool}\label{sec:tool}
\begin{figure}[t!]
    \centering
    \includegraphics[width=\columnwidth]{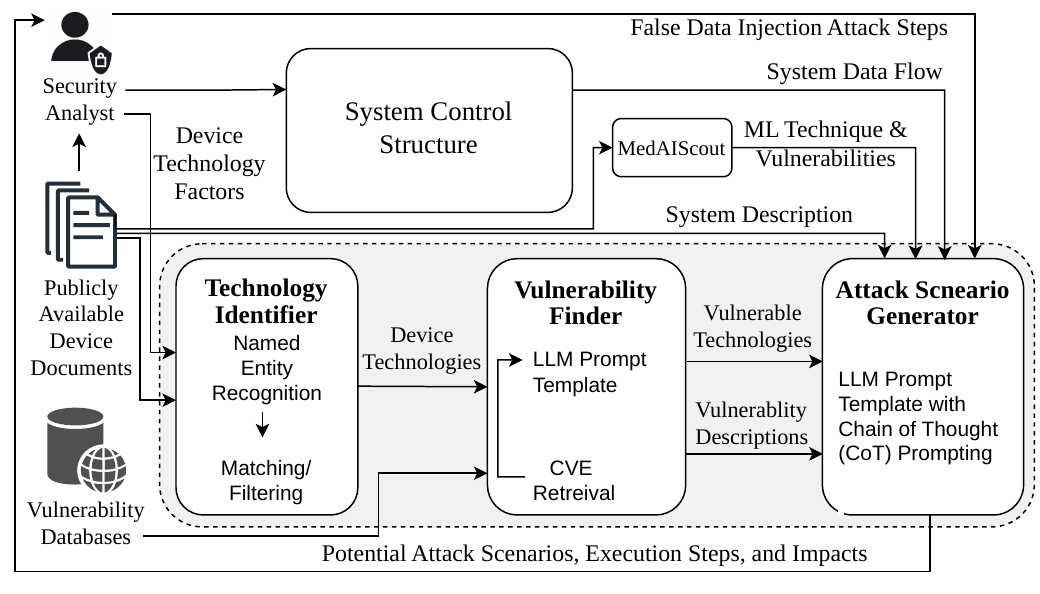}
    \caption{Workflow of \stpatool. Phases are highlighted in Bold.}
    \vspace{-1em}
    \label{fig:workflow}
\end{figure}

\stpatool is an automated tool that assists security analysts by addressing the aforementioned challenges in STPA-Sec analysis of AI/ML-enabled medical devices. 
Fig. \ref{fig:workflow} provides an overview of the workflow of \stpatool. For a target AI/ML-enabled medical device, \stpatool takes as input: (1) its system control structure, including the device's hardware and software components, the peripheral devices connected to the ML engine, and the control- and data-flow paths connecting all of these components; (2) the system description which includes the functionality and type of data processed by the device; (3) the ML technique employed by the device along with the associated vulnerabilities; and (4) the common adversarial steps and known types of technological attack vectors in false data injection attacks. A security analyst can obtain the (1) and (2) either from the device documentation or by requesting the device manufacturer. (3) can be inferred using tools such as MedAIScout~\cite{dharmalingam2024medaiscout}, the common adversarial steps and vulnerability information ((4)) are known from the MITRE ATT\&CK framework and derived from publicly accessible vulnerability databases, respectively. \stpatool{} uses this information along with available device documents to \textit{automatically} generate scenarios in which an adversary leverages known vulnerabilities in system components to manipulate data or inject false inputs into the ML engine during inference. For this, \stpatool{} leverages LLM reasoning, but avoids naive prompting, which typically yields generic threat descriptions without traceability to specific system components and vulnerabilities. Instead, \stpatool{} enforces the following analytical workflow: 
\\
First, \textbf{(1) Technology Identifier} finds all the technologies associated with each compatible peripheral device using NLP techniques
(\S\ref{sec:tech_identify}); 
Next, \textbf{(2) Vulnerability Finder} 
identifies known vulnerabilities in each architectural layer of the medical device and the peripheral devices by interfacing with CVE databases, and then filtering the relevant CVE records using LLMs (\S\ref{sec:vul_identify});
Finally, \textbf{(3) Attack Scenario Generator} generates attack steps to outline how the identified vulnerabilities could be exploited post-deployment to execute known attacks on the ML model, using a structured prompt that contains a description of known ML attacks, CVE description, and the technology layer of the peripheral device where the vulnerability was found. These scenarios outline all possible ways an adversary can inject malicious data into the ML engine and compromise the safety of the patient (\S\ref{sec:causal-scenario}). 

This structured encoding of threat modeling logic into the prompts yields two benefits: (i) it significantly reduces hallucinations in LLM outputs by constraining reasoning to system-specific components and known vulnerabilities; and (ii) it provides manufacturers with a holistic view of security risk.  
Using \stpatool{} is straightforward for security analysts. They would 
plug in their own LLM API keys and run \stpatool{} locally.

\subsection{Technology Identifier}\label{sec:tech_identify}



The Technology Identifier uses NLP- and LLM-based methods to extract information about the technologies used by all the system components mentioned in the system control structure, from the device documentation available on the FDA website~\cite{fdaml} and user guides prepared by the manufacturers. 
For instance, in the case of the BGMS shown in Fig~\ref{fig:BGMS_Example}, the technology identifier would identify Bluetooth and Wi-Fi as communication technologies used in the system.

To enable this automated identification of technologies, we use a combination of two named entity recognition (NER) models--GliNER~\cite{zaratiana-etal-2024-gliner} and GPT-4o-based NER~\cite{achiam2023gpt}, for their superior performance demonstrated in multiple domains. Given a list of entities (in our case, technology factors) as input, these NER models parse an input document written in natural language to extract keywords (in our case, technologies) that match the type of entities mentioned. The following entities are provided as input to the Technology Identifier: ``Communication Protocol'', ``Communication encryption'',	``Electromagnetic Susceptibility'', 	``Firmware'', ``Hardware'',	``Operating System'', and ``External Library and Data source''. This choice of technology factors is based on known attack vectors targeting AI/ML-enabled medical devices~\cite{yaqoob2019security}. 

To reduce hallucinations in the NER outputs, Technology Identifier performs an exact word match check of keywords extracted by each NER model and removes any keywords that do not appear in the device documentation. To ensure no technology information is missed, it then merges the filtered outputs from both models, removes duplicates, and produces a final list of all technologies used in the device.

\subsection{Vulnerability Finder}\label{sec:vul_identify}

\stpatool's Vulnerability Finder module compiles a list of known vulnerabilities associated with the technologies identified in the previous step. It automatically queries the MITRE Common Vulnerability Enumeration (CVE) database [39], a publicly accessible online vulnerability database, using the device technologies as search keywords and parses the results to extract the top-N (in this paper N is set to 10) most recent CVE records. 
By leveraging this continuously updated database, \stpatool covers all the newly discovered vulnerabilities. 

After collecting the CVE records, the Vulnerability Finder assesses their relevance to the target device by prompting an LLM 
with both the device description and the corresponding CVE description, instructing it to determine whether an adversary could exploit the vulnerability to inject false data into the device. The filtered set of CVEs would be finally used for generating attack scenarios.
Fig.~\ref{fig:cvefilter} shows the prompt used for filtering CVE records.

\begin{figure}[t]
    \centering
    \begin{tcolorbox}[enhanced,left=1mm, right=1mm,top=1mm, bottom=1mm]
\footnotesize
\texttt{``You are a security engineer. Decide if the vulnerability with (CVE No.: \{cve\} and Description: \{description\}) below could plausibly enable an attacker to INJECT, MODIFY, or SPOOF data used by an ML component in a medical device (\{device\}) and cause incorrect ML predictions. First, check if the technology factor \{factor\} is affected by this vulnerability \{cve\}. Proceed with the next steps only if the medical device (\{device\}) uses the \stpatool \{factor\}. If it doesn't, answer with NO. Otherwise, follow the next steps.}\\
\texttt{Answer strictly with:}\\
\texttt{YES  - if the \{cve\} plausibly allows data tampering/insertion/spoofing.}\\
\texttt{NO   - if it is only info disclosure/enum, UI redirect, DoS-only, or cannot reasonably lead to data being altered/inserted for the system.}\\
\texttt{Return ONLY the single token YES or NO.}\\[0.8em]
\texttt{Medical device: \{device\}}\\
\texttt{Technological factor: \{factor\}}\\
\texttt{CVE: \{cve\}}\\
\texttt{CNA: \{cna\}}\\
\texttt{CVE Description: \{description\}}\\[-0.2em]
\end{tcolorbox}
\captionsetup{justification=centering}
    \caption{\stpatool{} prompt template for CVE record selection}
    \vspace{-1em}
    \label{fig:cvefilter}
\end{figure}

\subsection{Attack Scenario Generator}\label{sec:causal-scenario}
The Attack Scenario Generator of \stpatool{} performs the most crucial step of STPA-Sec: identifying attack scenarios that could compromise patient safety. We leverage LLMs to automatically derive these scenarios from disclosed vulnerabilities in the system's underlying technologies. Because LLMs are trained on vast, cross-domain information publicly accessible on the web~\cite{openaidataset}, including information on cybersecurity and healthcare, they enable the Attack Scenario Generator to integrate interdisciplinary knowledge. As a result, LLMs provide technically grounded attack scenarios along with explicit analyses of their potential impact on patient 
safety.



A key challenge when using LLMs is prompt design--obtaining the ideal response while minimizing hallucinations. For \stpatool{}, an ideal response would include a detailed set of attack steps exploiting a technology vulnerability to perform inference-time data injection on a given ML technique, and the impact of the attack on the patient using the compromised device. To achieve this, we designed the prompt in Fig.~\ref{fig:attackstepgenerator}.

\begin{figure}[t]
    \centering
    \begin{tcolorbox}[enhanced, left=1mm, right=1mm, top=1mm, bottom=1mm]
\footnotesize
\texttt{Act as a security engineer who has the task of identifying the steps that an adversary follows to cause a security breach in an ML-enabled medical system. An ML-enabled system comprises an ML-enabled component collecting inputs from multiple peripheral devices and sending the predicted output to another peripheral device. A security breach is an event where a malicious attacker compromises the overall system’s confidentiality, integrity, or availability. You are given a system description, a data flow, an ML attack, a targeted input peripheral component, a targeted technology, and a known vulnerability in the input component. Give a list of steps to show how an adversary can exploit the vulnerability to mislead the ML-enabled component and how that affects the action of the output device on the patient.}\\
\texttt{Constraints:}\\
\texttt{- Do NOT include exploit code, commands, payloads}\\
\texttt{- Keep each step concise and high-level.}\\
\texttt{- You MUST use exactly these 5 step names, in this order:}\\
\texttt{\ \ 1) Reconnaissance}\\
\texttt{\ \ 2) Gaining access}\\
\texttt{\ \ 3) Privilege escalation}\\
\texttt{\ \ 4) Attack execution}\\
\texttt{\ \ 5) Impact}\\
\texttt{\textbf{System Description}: \{System description provided by the manufacturer in the device documentation\}}\\
\texttt{\textbf{Data flow}: \{This can be derived from the control structure\}}\\
\texttt{\textbf{ML attack}: \{The ML attack is identified using existing tools such as MedAIScout.\}}\\
\texttt{\textbf{Targeted technology}: \{One of the underlying technologies in the input device, as identified by the technology identifier (\S\ref{sec:tech_identify})\}}\\
\texttt{\textbf{Known vulnerability (CVE)}:} \{Description of the known vulnerability in the targeted technology, as retrieved from the CVE database in \S\ref{sec:vul_identify}\}\\
\texttt{- CVE: \{cve\_id\}}\\
\texttt{- CNA: \{cna\}}\\
\texttt{- Description: \{cve\_description\}}
\end{tcolorbox}
\captionsetup{justification=centering}
    \caption{\stpatool{} prompt template for generating attack steps}
    \vspace{-1em}
    \label{fig:attackstepgenerator}
\end{figure}

\vspace{0.5cm}
We observed that explicitly assigning the LLM the role of a security analyst improves the readability and relevance of the generated results. This is in line with prior work in this area~\cite{shanahan2023role,li2024camel,santu2023teler}. Similarly, providing the system data flow provides clarity to the LLM regarding the sequence of data transmission between different system components. We further guide the LLM to generate attack scenarios in a structured, stepwise manner using a fixed template that lists five attack stages applicable to all cyberattacks on ML-enabled medical devices~\cite{mitreattack}: ``(1) Reconnaissance'', ``(2) Gaining access'', ``(3) Privilege escalation'', ``(4) Attack execution'', and ``(5) Impact''.  
This chain-of-thought reasoning~\cite{wei2022chain} constraint and domain knowledge of attack stages improves completeness and consistency across generated scenarios, and reduces variability and hallucinations. We further mitigate hallucinations by empirically tuning the LLM temperature to a value that minimizes spurious outputs while preserving the model's reasoning capabilities. Hallucination rates are also very low in the latest models such as GPT-4o.

By running this prompt for technology used in the system and each vulnerability uncovered in it, \stpatool{} generates a comprehensive set of potential attack steps for the entire system. Device manufacturers or security analysts can then disregard those that have already been mitigated and develop design recommendations for the remaining ones. For example, in our BGMS case study, if a vulnerability is identified in the Bluetooth communication channel, the automatic attack scenario generator outlines a detailed sequence of steps--how an adversary could gain access to the communication link, manipulate transmitted data, and ultimately cause the system to deliver an incorrect insulin dose--along with the potential consequences of such an attack for the patient (e.g., death). 



\section{Experimental Setup}\label{evaluation}

\subsection{Case Studies}\label{case_study}
We applied \stpatool{} to five real-world FDA-cleared ML-enabled medical devices across different medical specialties: 

\begin{itemize}[leftmargin=*]
    \item \textbf{The d-Nav system~\cite{dnav}} is an ML-enabled app that recommends insulin doses to diabetic patients based on their blood glucose history and physiological data (e.g., meals, sleep). The patient or a caregiver configures the insulin pump to administer the recommended insulin dose to the patient.

\item \textbf{The ABMD software~\cite{abmd}} is an ML-enabled image processing software that measures bone mineral density from CT scans, and generates a report for the medical practitioner that help them make treatment plans.
\item \textbf{The IDx-DR v2.3~\cite{idx}} is an ML-enabled software system. It uses retinal images to automatically detect more-than-mild diabetic retinopathy in diabetic adults.

\item \textbf{The KIDScore D3~\cite{Kidscore}} is an ML-enabled software that provides fully automated analyses of developing embryos, which assist in selecting embryos for transfer, freezing, or further development.

\item \textbf{The Oxehealth Vital Signs~\cite{oxehealth}} is an ML-enabled software that utilizes video footage to non-invasively measure a resting patient's pulse rate and breathing rate.
\end{itemize}

We selected these devices for their wide adoption and for their diversity in clinical workflows and data modalities. For example, 
d-Nav is the only device among these that operates without direct involvement from a medical practitioner. Furthermore, ABMD (radiology) processes images, while Oxehealth Vital Signs (cardiovascular) analyzes video.

\subsection{Implementation}\label{implementation}
We used Python to develop all components of \stpatool{}. Furthermore, in the Attack Scenario Generator, we employed and evaluated three state-of-the-art large language models, viz., GPT-4, GPT-4o, and Llama 3, while we employed GPT-4o for the Vulnerability Finder. While \stpatool{} is compatible with any LLM, these models were selected for their advanced natural language processing capabilities and well-known effectiveness in generating contextually relevant outputs~\cite{achiam2023gpt,dubey2024llama}. 
%
We set the temperature of all the models to 0.7. We observed that raising the temperature above this value introduces unnecessary text without any useful information, thereby increasing the risk of hallucination, and lowering it restricts the model's ability to reason effectively and generate attack scenarios~\cite{lee2023mathematical}. 

\subsection{Evaluation Methods}

\textbf{Technology Identifier} is evaluated semi-automatically by searching the device document and manually checking the extracted list of technologies to indicate whether each extracted technological factor was actually present in the document (true positives) or not (false positives) or if any technological factors in the document were not extracted (false negatives). 

\textbf{Vulnerability Finder} is evaluated manually by reviewing the detailed descriptions of the filtered vulnerabilities and verifying whether they are relevant to false data injection attacks for the device under test.

\textbf{Attack Scenario Generator} is evaluated in two steps. In the first step, we select the best-performing LLM and among 3 LLMs, GPT-4, GPT-4o, and Llama 3, by manually evaluating the attack plans for a sample set of 5 scenarios (one vulnerability per device). In the second step, we proceed with the best-performing LLM and use an automated LLM-as-a-judge method to evaluate the correctness of all the generated attack scenarios. Each step is explained in more detail below: 

\begin{itemize}[leftmargin=*]
    \item \textbf{Manual Evaluation} assesses the generated attack scenarios from three perspectives: (1) \emph{Correctness}, measured as the fraction of completely correct steps among those generated. An attack step is considered correct if manual inspection confirms its practical feasibility, (2) \emph{Level of Detail}, measured by the extent to which the scenario specifies the attacker’s entry point, compromised system component, specific data manipulated or injected, and the resulting effect on the ML model’s output, and (3) \emph{Additional Information}, including mitigation strategies or security recommendations. These metrics provide medical device manufacturers with actionable insights, helping them prioritize defenses and design secure devices. 
    \item \textbf{LLM-as-a-judge Evaluation} prompts an independent LLM (more advanced than the LLM used for attack step generation) to assess each of the steps in the generated attack scenarios for contextual consistency with the system description, the referenced CVE, and relevance of false data injection. If any step is judged incorrect, we mark the entire attack plan as incorrect. We use GPT-o3 and Gemini-2.5-Pro as judges, due to their strong performance across diverse domains and their reasoning ability in judging other LLMs at a level comparable to humans~\cite{zheng2023judging}. We consider an attack plan to be correct only if both judges suggest that all the steps in the attack plan are correct. 
    
\end{itemize}

\section{Experimental Results}\label{results}
\setlength{\tabcolsep}{6pt}
\begin{table*}[t]
\centering
\begin{tabular}{l|c|c|c|c|c|c|c|c|c|c}
\hline
\multicolumn{1}{c|}{\multirow{2}{*}{\textbf{Devices}}} &
  \multicolumn{7}{c|}{\textbf{\begin{tabular}[c]{@{}c@{}}Technology Identification\end{tabular}}} &
  \multicolumn{2}{c|}{\textbf{\begin{tabular}[c]{@{}c@{}}Vulnerability \\ Identification\end{tabular}}} &
  \multicolumn{1}{c}{\multirow{2}{*}{\textbf{\begin{tabular}[c]{@{}c@{}}ASG Average Accuracy (\%)\\ (determined by LLM-as-a-judge)\end{tabular}}}} \\ \cline{2-10}
\multicolumn{1}{c|}{} &
  \multicolumn{1}{c|}{\textbf{CP}} &
  \multicolumn{1}{l|}{\textbf{ENCR}} &
  \multicolumn{1}{l|}{\textbf{EM}} &
  \multicolumn{1}{l|}{\textbf{FW}} &
  \multicolumn{1}{l|}{\textbf{HW}} &
  \multicolumn{1}{l|}{\textbf{OS}} &
  \multicolumn{1}{l|}{\textbf{EXT}} &
  \multicolumn{1}{c|}{\textbf{AutoCVE}} &
  \multicolumn{1}{c|}{\textbf{VerifiedCVE}} &
  \multicolumn{1}{c}{} \\ \hline
\multicolumn{1}{l|}{d-Nav}                 & {\color{green}\ding{51}}(2) & {\color{green}\ding{51}}(1) & {\color{black}-} & {\color{black}-} & {\color{black}-} & {\color{green}\ding{51}}(2) & {\color{green}\ding{51}}(5) & 9 & 4 & 100 \\\hline
\multicolumn{1}{l|}{ABMD}                  & {\color{black}-} & {\color{black}-} & {\color{black}-} & {\color{black}-} & {\color{black}-} & {\color{green}\ding{51}}(1) & {\color{green}\ding{51}}(1) & 8 & 6 & 100 \\\hline
\multicolumn{1}{l|}{IDx-DR v2.3}           & {\color{green}\ding{51}}(1) & {\color{black}-} & {\color{black}-} & {\color{black}-} & {\color{green}\ding{51}}(1) & {\color{green}\ding{51}}(3) & {\color{green}\ding{51}}(6) & 22 & 17 & 76.5 \\\hline
\multicolumn{1}{l|}{KIDScore D3}           & {\color{black}-} & {\color{black}-} & {\color{black}-} & {\color{black}-} & {\color{black}-} & {\color{black}-} & {\color{green}\ding{51}}(2) & 1 & 1 & 100 \\\hline
\multicolumn{1}{l|}{Oxehealth Vital Signs} & {\color{green}\ding{51}}(1) & {\color{black}-} & {\color{black}-} & {\color{black}-} & {\color{black}-} & {\color{black}-} & {\color{green}\ding{51}}(4) & 17 & 8 & 100 \\\hline
\end{tabular}%
\caption{Performance Evaluation of Different Modules in SAMD:
CP: Communication protocol, ENCR: Encryption technique used for securing communication, EM: Susceptibility to electromagnetic radiation, FW: Firmware, HW: Hardware, OS: Operating System, EXT: External libraries or data sources used; {\color{green}\ding{51}}: Present in documentation and correctly identified by SAMD, {\color{black}-}: Not present in device documentation; AutoCVE: No. of CVEs filtered as relevant to false data injection by SAMD, VerifiedCVE: No. of CVEs manually verified as relevant to false data injection, ASG: Attack Scenario Generation}
\vspace{-1em}
\label{tab:master-results}
\end{table*}
\noindent\textbf{Efficiency in Technology Identification. } 
Across all 5 devices, a total of 30 different technologies were specified in the device documentation and user guides. \stpatool{} was able to identify all of them (refer to Table~\ref{tab:master-results} for device-wise numbers), resulting in an overall identification accuracy of $100\%$. 

However, it should be noted that the publicly available device documentation that we used for testing \stpatool did not contain all the device technology information, as many of them are kept confidential by the manufacturers (indicated by {\color{black}-} in Table~\ref{tab:master-results}). This results in missed technology information in \stpatool and less comprehensive attack scenario identification. However, this is not a drawback of \stpatool, as it could identify all technologies mentioned in its input documents. Device manufacturers and security analysts would likely have access to more detailed device specifications that they can provide to \stpatool for a more thorough analysis.
\noindent\textbf{Ability to discover relevant vulnerabilities. } For every technology identified for each device, \stpatool{} was able to automatically extract the top 10 recent CVE records from the MITRE CVE database, if they existed. Across all 5 devices, \stpatool extracted a total of 165 vulnerabilities, out of which it selected 57 vulnerabilities as exploitable for false data injection. 
During manual verification, we found 36 out of 165 vulnerabilities to be truly exploitable for false data injection attacks (true positives). Device-wise number of vulnerabilities identified by SAMD (AutoCVE) vs. those identified manually (VerifiedCVE) are shown in Table~\ref{tab:master-results}. 
Upon closer inspection, we found that out of the 36 CVEs, 35 are recent (discovered in 2025), 14 of them are not yet patched, and among the unpatched vulnerabilities, 3 are marked as severe/critical by MITRE (i.e., they are extremely highly exploitable and must be prioritized for mitigation). 

We also evaluated whether \stpatool missed any relevant CVEs from the initial top-10 retrieved CVEs (false negatives). Because the set of remaining CVEs was large (108 CVEs), we manually reviewed a random 20\% sample set (22 cases). None of them were related to false data injection, indicating a 100\% recall in the sampled set. These results highlight \stpatool{}'s efficacy in identifying relevant and critical vulnerabilities with a precision of $63.2\%$. 
\noindent\textbf{Performance of Attack Scenario Generator.} Using the technology and vulnerability information extracted in the previous steps, the ML technique and vulnerability information obtained using MedAIScout~\cite{dharmalingam2024medaiscout}, and the system description obtained from the device documentation and user guides, we crafted prompts for the LLM-based attack scenario generator using the template shown in \S~\ref{sec:causal-scenario}. 
MedAIScout identified the following ML techniques and potential vulnerabilities: \textbf{d-Nav} uses an ML-based insulin dose calculator processing time-series data and may be vulnerable to model inversion attacks requiring false data injection~\cite{fredrikson2015model}; \textbf{ ABMD} uses ML-based CT image processing and may be susceptible to GAN-based malicious image tampering~\cite{mirsky2019ct}; \textbf{IDx-DR v2.3} employs DNN-based retinal image analysis and may be exposed to adversarial exposure attacks~\cite{cheng2024adversarial}; \textbf{ KIDScore D3} uses predictive algorithms with automated cell tracking and may be vulnerable to projected gradient descent attacks~\cite{ghaffari2022adversarial}; and \textbf{Oxehealth Vital Signs} uses ML-based image analysis that may be affected by adversarial perturbation attacks~\cite{li2018adversarial}.

We had a total of 36 prompts for 36 relevant technology vulnerabilities identified across the 5 devices, while the system description, ML technique, and ML vulnerability of each device were kept constant. We perform two steps. 

\textit{Step 1: Selecting the best-performing LLM by manual evaluation: } For this, we execute the attack scenario generator for five \{technology, vulnerability\} combinations (1 per device), as shown in Table~\ref{tab:details}. We repeat this step for three LLMs--GPT-4, GPT-4o and Llama 3. The results are as follows. 

\begin{figure}[t]
        \centering
        \includegraphics[width=\linewidth]{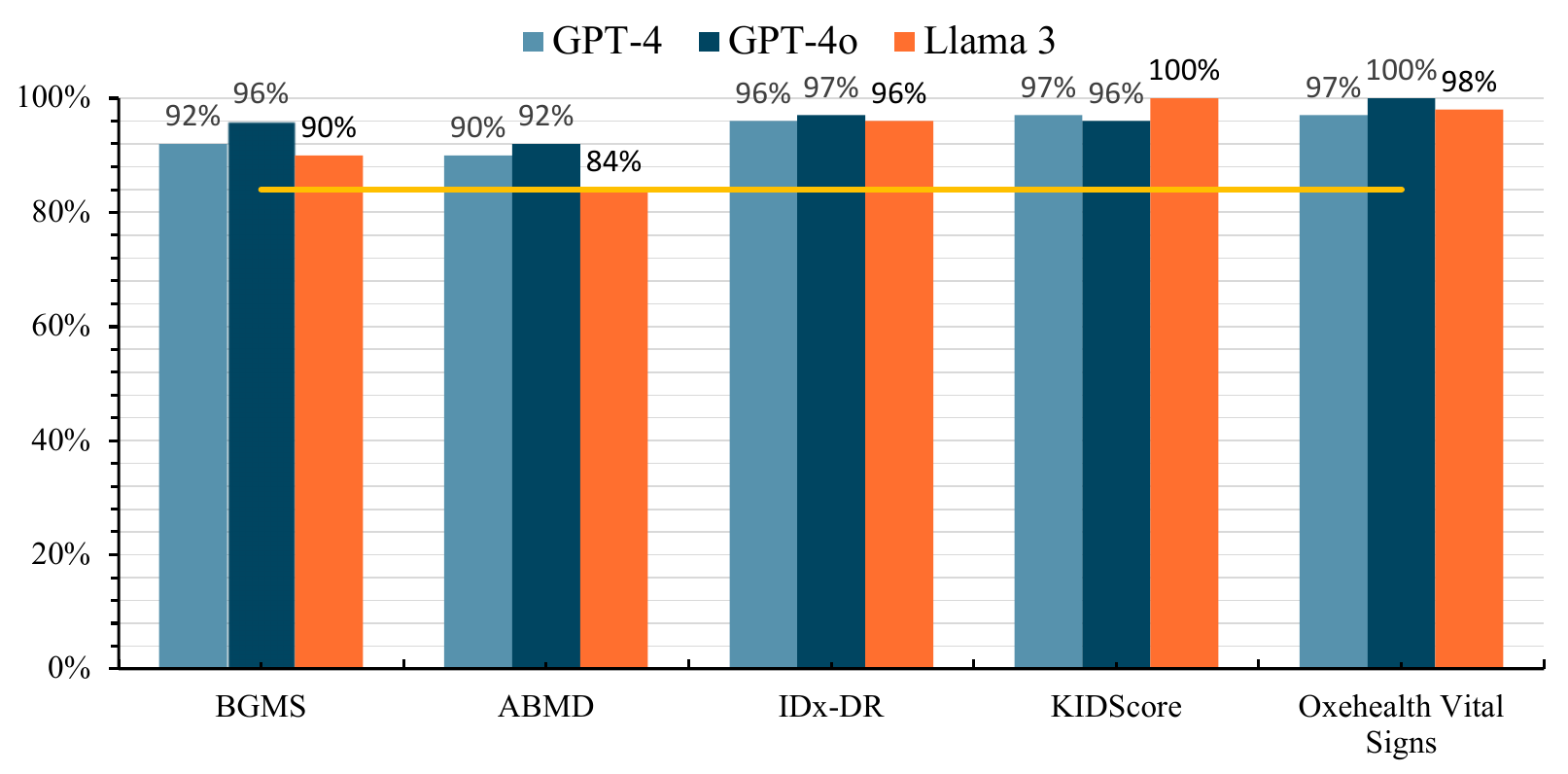}
        \caption{Comparison of output correctness among three LLMs across different case studies (The yellow line indicates the lowest among all).}
        \vspace{-1em}
        \label{fig:LLM_Chart}
    \end{figure}

\textbf{Correctness. }Figure~\ref{fig:LLM_Chart} shows the correctness of the attack steps for the 3 LLMs across the 5 scenarios. GPT-4o consistently produced more correct attack steps ($96.2\%$), as compared to both GPT-4 ($94.4\%$) and Llama 3 ($93.6\%$). 

\begin{table*}[h]
        \centering
        \begin{tabular}{p{1cm}|p{2cm}|p{1cm}|p{5.5cm}|p{6cm}}
        \hline
           \textbf{Device}  & \textbf{Technology}  & \textbf{CVE} & \textbf{Level of details regarding attack strategy} & \textbf{Additional information provided}\\
           \hline
           d-Nav &Wi-Fi Communication & CVE-2023-35836 & GPT-4o provides details on how to gain access and eavesdrop on the communication. & GPT-4o provides a summary of the impact on C.I.A. and patient safety. It also provides a set of security recommendations specific to the system and the vulnerable point. \\
           \hline
           ABMD & Ethernet Communication & CVE-2023-1670 & GPT-4o breaks down each step into objective and action. & Both GPT-4 and GPT-4o provide the impact on CIA. Llama 3 highlights the importance of securing the entire system at the end. \\
           \hline
           IDx-DR v2.3 & Linux Operating System & CVE-2019-0708 & GPT-4o breaks down each step into objectives and Tools/Techniques. It comes with more steps, which Llama3 follows closely.  GPT-4o also breaks down the impact on the patient into two scenarios: False Positive, and False Negative. & GPT-4o comes up with a way to avoid detection by security systems and how to cover the attack by erasing the traces of manipulation. GPT-4 and Llama 3 mention the impact on C.I.A. \\
           \hline
           KIDScore D3 & Human Interaction& -- &Llama 3 comes up with the most detailed steps, GPT-4o follows closely. & Llama 3 provides an additional step to evade detection.\\
           \hline
           Oxehealth Vital Signs & Patient monitoring camera & CVE-2024-46959 & GPT-4o, and Llama 3 come with more details on each step. & GPT-4o provides suggestions for evading defenses. \\
           \hline
        \end{tabular}
\captionsetup{justification=centering}
\caption{Examples of varying levels of detail provided by different LLMs across the five case studies.}
\vspace{-1em}
\label{tab:details}
\end{table*}

\textbf{Level of detail. }Across the five scenarios, GPT-4o provided the most detailed steps as compared to the other models. For most vulnerabilities, it clearly defined the objective of each step and the required actions to be taken. We present key examples in Table \ref{tab:details} to illustrate these claims.

\textbf{Additional information. }Across the five scenarios, GPT-4o consistently provided detailed analyses of attacks' impact on Confidentiality, Integrity, and Availability (CIA) and, in some cases, proposed mitigation strategies. Llama 3 and GPT-4 offered partial insights, particularly regarding patient impact. We present key examples in Table \ref{tab:details}.

\textit{Step 2: Evaluating the attack scenarios generated by the best-performing LLM: }Since the previous step showed that
GPT-4o surpasses the two other LLMs in terms of correctness, level of details, and additional information, we integrate GPT-4o in the Attack Scenario Generator for the remaining experiments. Across the 5 selected devices, we executed the Attack Scenario Generator for 36 unique \{technology, vulnerability\} combinations. Table~\ref{tab:master-results} shows the accuracy (in \%) of the generated attack scenarios, as determined jointly by the two LLM judges. An attack scenario is considered accurate if \emph{all} its attack steps are correct.
Figure~\ref{fig:attackdemo} shows a representative output of the attack-scenario generator for IDx-DR, for CVE-2025-5307, identified in a compatible DICOM server (peripheral device).
\emph {On average, \stpatool{} shows an accuracy of $95.3\%$}. 

\begin{figure}[t]
\centering
\begin{tcolorbox}[enhanced,left=1mm, right=1mm,top=1mm, bottom=1mm]
\footnotesize
\texttt{
\noindent\textbf{Reconnaissance:} The adversary gathers information about the IDx-DR system, focusing on the data flow and identifying the Topcon NW400 fundus camera and the DICOM protocol as key components. They also research the CVE-2025-5307 vulnerability in the Sante DICOM Viewer Pro, which is used in the system.\\
\textbf{Gaining access: }The adversary gains access to the local network where the Sante DICOM Viewer Pro is installed, potentially through phishing or exploiting weak network security measures, allowing them to interact with the DICOM data.\\
\textbf{Privilege escalation:} The adversary exploits the CVE-2025-5307 vulnerability in the Sante DICOM Viewer Pro to execute arbitrary code, gaining higher privileges on the system to manipulate DICOM data.\\
\textbf{Attack execution:} Using the elevated privileges, the adversary alters the camera exposure settings in the DICOM data, creating adversarial examples that mislead the IDx-DR's DNNs into incorrectly grading the severity of diabetic retinopathy.\\
\textbf{Impact:} The manipulated results are sent back to the medical practitioner, potentially leading to incorrect treatment decisions for the patient, affecting their health outcomes due to misdiagnosis or inappropriate treatment plans.
}
\end{tcolorbox}
    \caption{Example output of Attack Step Generator for IDx-DR. Based on this output, the ML model designer may harden the model against image manipulation, enforce DICOM metadata authentication, require secure local networks, or restrict the use of Sante DICOM Viewer Pro with their model.}
    \vspace{-1em}
    \label{fig:attackdemo}
\end{figure}

\textbf{Efficiency gain via automation.} The total runtime of \stpatool{} ranged between 8.74s (for KidScore D3) to 191.64s (for IDx-DR v2.3). The time required to identify and analyze security vulnerabilities for a given device scales linearly with the number of vulnerabilities present in its peripheral components. We empirically observed that while manual analysis requires, on average, 2 minutes per CVE to assess relevance and 10 minutes per relevant CVE to outline the attack steps, \stpatool{} requires 5 seconds per CVE for relevance classification and 12 seconds per relevant CVE for attack-step generation. Therefore, for a device having $x$ CVE records associated with its peripheral devices, of which $y$ are relevant to inference-time false data injection, manual analysis would take $\approx(2x+10y)$ \textit{minutes}, while \stpatool{} would require $\approx(5x+12y)$ \textit{seconds}. For IDx-DR (54 CVEs, 22 relevant), manual analysis would take 5.5 hours, whereas \stpatool{} took approximately 3 minutes.
This high accuracy and speed demonstrate that \stpatool{} efficiently and reliably produces correct, detailed attack scenarios across diverse devices, technologies, and vulnerabilities. 

\section{Discussion}\label{discussion}
\textbf{Utility of \stpatool{}.}
\stpatool is designed to provide actionable insights to various stakeholders throughout a device's life cycle, i.e., starting from the design stage through development, deployment, and operation. \stpatool{} helps device manufacturers and independent security analysts to proactively identify and mitigate vulnerabilities in AI/ML-enabled medical devices, and ensure patient safety.
\begin{enumerate}[leftmargin=*]
    \item \textbf{Device manufacturers} can use \stpatool during the design phase to identify vulnerabilities in system components and incorporate appropriate mitigations at an early stage. The tool can also help them determine which peripheral components should be supported based on their security posture. Post-deployment, manufacturers can use \stpatool{} to continuously assess the device's exposure to newly discovered vulnerabilities, strengthen security measures, and issue security advisories for their end users.
    \item \textbf{Independent security analysts} can use \stpatool to investigate attack paths, assess their impact, recommend mitigations, and translate technical findings into actionable security guidelines for non-technical stakeholders such as patients, clinicians, and IT personnel of healthcare organizations.
\end{enumerate}

\textbf{Limitations of \stpatool}

Although LLMs have shown significant potential in generating realistic attack scenarios and validating them, we  still recommend a final human review, considering the high safety-criticality of the healthcare domain. \stpatool is intended to support, and not replace human security analysts (as of yet).

\section{Conclusion}\label{conclusion}
 

This paper introduces \stpatool{}, an automated tool that assists in systematically performing STPA-Sec on ML-enabled medical devices throughout the device lifecycle. \stpatool{} will enable medical device manufacturers, security analysts, and healthcare organizations to make secure design choices and perform continuous security risk assessments. Leveraging LLMs, \stpatool{} automates the identification of potential adversarial false data injection scenarios that could cause an ML model mispredict a patient's condition or generate an incorrect treatment plan. We evaluated \stpatool{} using five real-world FDA-cleared medical devices from different medical specialties. The results show that \stpatool{} effectively identifies a comprehensive set of potential attack scenarios and their associated patient safety impacts, with an average accuracy of $\approx95\%$, and the highest time taken being 191.64s. Among different LLMs, GPT-4o provides more detailed steps and additional insights on potential mitigation approaches and security recommendations. 
\FloatBarrier


\bibliographystyle{IEEEtran}
\bibliography{References}









\end{document}